# Rumor Detection on Social Media: Datasets, Methods and Opportunities

Quanzhi Li, Qiong Zhang, Luo Si, Yingchi Liu
Alibaba Group, US
Bellevue, WA, USA
{quanzhi.li, qz.zhang, luo.si, yingchi.liu}@alibaba-inc.com

**Abstract**

Social media platforms have been used for information and news gathering, and they are very valuable in many applications. However, they also lead to the spreading of rumors and fake news. Many efforts have been taken to detect and debunk rumors on social media by analyzing their content and social context using machine learning techniques. This paper gives an overview of the recent studies in the rumor detection field. It provides a comprehensive list of datasets used for rumor detection, and reviews the important studies based on what types of information they exploit and the approaches they take. And more importantly, we also present several new directions for future research.

## 1 Introduction

Rumors sometimes may spread very quickly over social media platforms, and rumor detection has gained great interest in both academia and industry recently. Government authorities and social media platforms are also taking efforts to defeat the negative impacts of rumors. In the following sub sections, we first introduce the rumor detection definition, the problem statement, and user stance, an important concept for the rest of this paper.

### 1.1 Rumor Detection

Different publications may have different definitions for rumor. It is hard to do a head-to-head comparison between existing methods due to the lack of consistency. In this survey, a rumor is defined as a statement whose truth value is *true, unverified* or *false* (Qazvinian et al., 2011). When a rumor's veracity value is *false*, some studies call it *"false rumor"* or *"fake news"*. However, many previous studies give *"fake news"* a stricter definition: fake news is a news article published by a news outlet that is intentionally and verifiably false (Vosoughi et al., 2018; Shu et al., 2017a; Cao et al., 2018). The focus of this study is rumor on social media, not fake news. There are also different definitions for *rumor detection*. In some studies, rumor detection is defined as determining if a story or online post is a rumor or non-rumor (i.e. a real story, a news article), and the task of determining the veracity of a rumor (*true, false* or *unverified*) is defined as rumor verification (Zubiaga et al., 2016; Kochkina et al., 2018). But in this survey paper, as well as in (Ma et al., 2016; Cao et al., 2018; Shu et al, 2017; Zhou et al., 2018), *rumor detection* is defined as determining the veracity value of a rumor. This means it is the same as *rumor verification* defined in some other studies.

### 1.2 Problem Statement

The rumor detection problem is defined as follow: A story $x$ is defined as a set of $n$ pieces of related messages $M = \{m_1, m_2, ..., m_n\}$. $m_1$ is the source message (post) that initiated the message chain, which could be a tree-structure having multiple branches. For each message $m_i$, it has attributes representing its content, such as text and image. Each message is also associated with a user who posted it. The user also has a set of attributes, including name, description, avatar image, past posts, etc. The rumor detection task is then defined as: Given a story $x$ with its message set $M$ and user set $U$, the rumor detection task aims to determine whether this story is *true, false* or *unverified* (or just *true* or *false* for datasets having just two labels). This definition formulates the rumor detection task as a veracity classification task. The definition is the same as the definition used in many studies (Cao et al., 2018; Shu et al, 2017b; Ma et al., 2016; Zhou et al., 2018).

### 1.3 User Stance

User responses to a source post (the first message) have been exploited in some rumor detection models. Most studies use four stance categories:

| Dataset | Total rumors (claims) | Text | User info | Time stamp | Propagation info | Platform | Description |
|---|---|---|---|---|---|---|---|
| PHEME-R | 330 | y | y | y | y | Twitter | Tweets from [Zubiaga et al., 2016] |
| PHEME | 6425 | y | y | y | y | Twitter | Tweets from [Kochkina et al., 2018] |
| Ma-Twitter | 992 | y | y | y | | Twitter | Tweets from [Ma et al., 2016] |
| Ma-Weibo | 4,664 | y | y | y | | Weibo | Weibo data from [Ma et al., 2016] |
| Twitter15 | 1,490 | y | y | y | y | Twitter | Tweets from [Liu et al., 2015; Ma et al.,2016] |
| Twitter16 | 818 | y | y | y | y | Twitter | Tweets from [Ma et al., 2017b] |
| BuzzFeedNews | 2,282 | y | | | | Facebook | Facebook data from [Silverman et al., 2016] |
| SemEval19 | 325 | y | y | y | y | Twitter, Reddit | SemEval 2019 Task 7 data set. |
| Kaggle Emergent | 2145 | y | | | | Twitter, Facebook | Kaggle rumors based on Emergent.info |
| Kaggle Snopes | 16.9K | y | | | | Twitter, Facebook | Kaggle rumors based on Snopes.com |
| Facebook Hoax | 15.5K | y | y | y | | Facebook | Facebook data from [Tacchini et al., 2017] |
| Kaggle PolitiFact | 2923 | y | y | y | y | Twitter | Kaggle rumors based on PolitiFact |
| FakeNewsNet | 23,196 | y | y | y | y | Twitter | Dataset from [Shu et al., 2019], enhanced from PolitiFact and GossipCop |

Table 1: Datasets for rumor detection and their properties

*supporting*, *denying*, *querying* and *commenting*. Some studies have explicitly used stance information in their rumor detection model, and have shown big performance improvement (Liu et al., 2015; Enayet and El-Beltagy, 2017; Ma et al., 2018a; Kochkina et al., 2018), including the two systems, (Enayet and El-Beltagy, 2017) and (Li et al., 2019a), that were ranked No. 1 in SemEval 2017 and SemEval 2019 rumor detection tasks, respectively. Stance detection is not the focus of this paper, but stance information has been used explicitly or implicitly in many rumor detection models, and in the next section we will also discuss some multi-task learning approaches that jointly learn stance detection and rumor detection models.

In the following sections, we will *1.* introduce a comprehensive list of datasets for rumor detection, *2.* discuss the research efforts categorized by the information and approaches they use, and *3.* present several directions for future research

## 2 Datasets and Evaluation Metrics

### 2.1 Datasets

Datasets could vary depending on what platforms the data are collected from, what types of contents are included, whether propagation information is recorded, and so on. Table 1 lists the datasets for rumor detection. There are also other datasets for fake news detection. Because this paper focuses on rumor detection on social media, and those datasets are only for fake news detection and do not have social context information (e.g. user responses, user data, and propagation information), so we did not list them here. The data of datasets in Table 1 are collected from four social media platforms: Twitter, Facebook, Reddit and Weibo. Weibo is a Chinese social media platform with over 400 million users, and it is very similar to Twitter. More than half of these datasets have three veracity labels: *true, false* and *unverified*. Others have only two labels: *true* and *false*. Among these datasets, PHEME-R has been used by SemEval 2017 rumor detection task and SemEval19 has been used by SemEval 2019 rumor detection task (Gorrell et al., 2019). The dataset links are listed below:

- *PHEME-R*: https://figshare.com/articles/PHEME_rumour_scheme_dataset_journalism_use_case/2068650
- *PHEME*: https://figshare.com/articles/PHEME_dataset_for_Rumour_Detection_and_Veracity_Classification/6392078
- *Ma-Twitter*: http://alt.qcri.org/~wgao/data/rumdect.zip
- *Ma-Weibo*: http://alt.qcri.org/~wgao/data/rumdect.zip
- *Twitter15*: https://www.dropbox.com/s/7ewzdrbelpmrnxu/rumdetect2017.zip?dl=0
- *Twitter16*: https://www.dropbox.com/s/7ewzdrbelpmrnxu/rumdetect2017.zip?dl=0
- *BuzzFeedNews*: https://github.com/BuzzFeedNews/2016-10-fac\ebook-fact-check

- *SemEval19*: https://competitions.codalab.org/competitions/19938#learn_the_details-overview
- *Kaggle Emergent*: https://www.kaggle.com/arminehn/rumor-citation
- *Kaggle Snopes*: https://www.kaggle.com/arminehn/rumor-citation
- *Facebook Hoax*: https://github.com/gabll/some-like-it-hoax/tree/master/dataset
- *Kaggle PolitiFact*: https://www.kaggle.com/arminehn/rumor-citation
- *FakeNewsNet*: https://github.com/KaiDMML/FakeNewsNet

## 2.2 Evaluation Metrics

Most existing approaches consider rumor detection as a classification problem. Usually it is either a binary (*true* or *false*) or a multi-class (*true, false* or *unverified*) classification problem. The evaluation metrics used the most are precision, recall, *F1* and accuracy measures. Because some datasets are skewed, Macro *F1* measure will provide a better view on the algorithm performance over all classes. Here we briefly describe them. For each class *C*, we calculate its precision (*p*), recall (*r*) and *F1* score as follow:

$$p = \frac{\text{no. of rumors predicetd as C correctly}}{\text{no. of rumors predicted as C}} \quad (1)$$

$$r = \frac{\text{no. of rumors predicetd as C correctly}}{\text{no. of rumors annotated as C}} \quad (2)$$

$$F1 = \frac{2*p*r}{p+r} \quad (3)$$

Consider all the classes together, then the Macro *F1* score is:

$$Macro\ F1 = \frac{1}{n} \sum_{i=1}^{n} F1_i \quad (4)$$

where *n* is the number of classes, and $F1_i$ is the score for class *i*. The overall accuracy for all the rumor types is:

$$accuracy = \frac{\text{no. of rumors predicetd correctly}}{\text{no. of rumors}} \quad (5)$$

## 3 Features and Approaches

In this section, we review previous studies based on the type of information they exploited in their models. The information for rumor detection can be categorized from several information dimensions: content, user, propagation path, and network. We will also give a brief overview for studies employing multi-task learning for stance detection and rumor detection, and introduce the contests for rumor detection. Table 2 presents the studies and their related information. From this table we can see that most studies have exploited text content, user information and propagation path. A few of them also explicitly incorporate user stance in their models. It also shows that almost all the most recent studies utilized neural networks in their models. Due to the space limitation, we just describe the representative studies in this paper.

### 3.1 Approaches Using Content Information

**Textual Content**. Text content is utilized by almost all the previous studies on rumor detection. It includes the source post and all user replies. According to deception style theory, the content style of deceptive information that aims to deceive readers should be somewhat different from that of the truth, e.g., using exaggerated expressions or strong emotions. And from user response text, we can also explore stance and opinion of users towards rumors.

Generally, text features can be grouped into attribute-based or structure-based features (Zhou and Zafarani, 2018). Attribute-based features include quantity (word, noun, verb, phrase, etc.), uncertainty (number of question mark, quantifiers, tentative terms, modal terms), subjectivity (percentage of subjective verbs, imperative commands), sentiment (positive/negative words, exclamation marks), diversity (unique content words, unique function words), and readability. Structure-based features include lexicon, syntax, semantic and discourse information, such as part-of-speech taggers, context-free grammar, and Linguistic Inquiry and Word Count (LIWC).

An early study from (Castillo et al., 2011) uses many text features in their model, such as the fraction of tweets with hashtags. These features and other additional text features are also used in other studies (Liu et al., 2015; Enayet and El-Beltagy, 2017; Li et al., 2019a; Ma et al., 2017; Li et al., 2019b). Kwon et al. (2013) also use LIWC dictionaries. Chua and Banerjee (2016) analyzed six categories of features: comprehensibility, sentiment, time orientation, quantitative details, writing style, and topic. Some important features reported were: negation words, past, present, future POS in the tweets, discrepancy, sweat and exclusion features. Textual content plays an important role in rumor detection, but most studies show that just utilizing text content is not enough.

**Visual Content:** Visual features (images or videos) have been shown to be an important indicator for rumor detection (Jin et al., 2017a; Jin et al., 2017b; Shu et al., 2017a; Wang et al., 2018). Rumors exploit the individual vulnerabilities and often use sensational or fake images to provoke emotional user responses. There are two visual feature types: statistical features and content features. Statistical features include image/video count, image ratio, etc. (Gupta et al., 2013; Jin et al., 2017a; Jin et al., 2017b; Shu et al., 2017a; Liu et al., 2015; Li et al., 2019a; Li et al., 2019b; Shu et al., 2017). Visual content features include clarity score, coherence score, diversity score, similarity distribution histogram, etc. (Wang et al., 2018; Shu et al., 2017). Jin et al. (2017a; 2017b) use various visual content and statistical features for rumor detection. Wang et al. (2018) employ a multi-modal feature extractor to extract the textual and visual features from posts, and then the textual feature representation and visual feature representation are concatenated together to form the multi-modal feature representation.

| Study | Information Used | | | | | | Approach |
|---|---|---|---|---|---|---|---|
| | Text | Visual | User | Propagation | Network | Explicitly using user stance | |
| [Castillo et al., 2011] | y | | y | y | | | DT |
| [Chang et al., 2016] | y | y | y | | | | Clustering |
| [Chen et al., 2016] | y | | y | y | | y | Anomaly detection, KNN |
| [Chua and Banerjee, 2016] | y | | | | | | LR |
| [Enayet and El-Beltagy, 2017] | y | | y | | | y | SVM |
| [Giasemidis et al., 2016] | y | | y | | y | | DT |
| [Gupta et al., 2012] | y | | y | | y | | Graph |
| [Gupta et al., 2013] | | y | y | | y | | DT, Graph |
| [Jin et al., 2016] | y | | y | | y | y | Graph, LDA |
| [Kwon et al., 2013] | y | | y | y | | | SVM, RF, LR |
| [Kwon et al., 2017] | y | | y | y | | | SpikeM |
| [Li et al., 2016] | y | | y | y | | | SVM |
| [Li et al., 2019] | y | y | y | y | | y | Deep NN, LSTM |
| [Liu et al., 2015] | y | y | y | | | y | SVM |
| [Liu and Wu, 2018] | y | | | y | | | CNN, RNN |
| [Ma et al., 2017] | y | | | y | | | NN |
| [Ma et al., 2015] | y | | | | | | SVM, RF, DT |
| [Ma et al., 2018a] | y | | | y | | | LSTM, multi-task |
| [Ma et al., 2018b] | y | | | y | | | Recursive NN |
| [Qin et al., 2016] | y | | | | | | SVM |
| [Shu et al., 2017b] | y | | y | | y | | NN |
| [Vosoughi, 2015] | y | | y | y | | | HMM |
| [Wang and Terano, 2015] | y | | y | | y | y | Graph |
| [Wang et al., 2018] | y | y | | | | | CNN, Adversarial NN |
| [Wu et al., 2015] | y | | y | y | | | SVM |
| [Yang et al., 2012] | y | | y | | | | SVM |
| [Yang et al., 2015] | y | | y | | y | | Graph |
| [Yang et al., 2018] | y | | | y | | | CNN |
| [Zhang et al., 2018] | y | | y | | y | | RNN |

Table 2: Previous studies, used information, and methods. Note: SVM - support vector machine, RF - random forest, DT- decision tree, LR – logistic regression, KNN – k nearest neighbor, NN – neural network, HMM – hidden Markov model.

### 3.2 Approaches Exploiting User Information

Users engage in rumor dissemination in multiple ways, such as sharing, liking, forwarding and reviewing. Many previous studies have shown that user credibility information is very important in rumor verification (Castillo et al., 2011; Yang et al., 2012; Gupta et al., 2012; Liu et al., 2015; Vosoughi, 2015; Shu et al., 2017b; Zhang et al., 2018; Liu and Wu, 2018; Li et al., 2019a; Li et al., 2019b). Based on 421 false rumors and 1.47 million related tweets, Li et al. (2016) study various semantic aspects of false rumors, and analyze their spread and user characteristics. Some findings are: when people do not have clarity about the veracity of a rumor, they usually just spread it without adding their opinions; credible users are less likely to support rumors, while low credibility accounts provide the most support; in terms of supporting or debunking a rumor, credible users are much more stringent, and hence a more trustworthy source than their corresponding counterparts.

Hand-crafted user features like registration age of users, number of followers, the number of posts the user had authored, and the like, are leveraged along with other textual and propagation features in Castillo et al. (2011) and other studies (Liu et al., 2015; Enayet and El-Beltagy, 2017; Li et al., 2019a; Li et al., 2019b). Liu and Wu (2018) construct user representations using network embedding approaches on the social network graph. There has been evidence that lots of rumors come from either fake news websites or hyper-partisan websites (Silverman, 2016; Li et al., 2016; Liu et al., 2015).

## 3.3 Approaches Based on Propagation Path and Network

Rumors spread through social media in the form of shares and re-shares of the source post and shared posts, resulting in a diffusion cascade or tree. The path of re-shares and other propagation dynamics are utilized for rumor detection. We group current studies into (1) cascade-based rumor detection techniques, which take direct advantage of rumor propagation paths, and (2) network-based detection methods, which construct a flexible network from cascades, from which rumors are indirectly detected.

**Propagation-based:** When using cascades to detect rumors, one either distinguishes them by computing the similarity of its cascade to that of other true/false rumors, or by generating a cascade representation that facilitates distinguishing false and true rumors. Ma et al. (2018b) construct a tree-structured neural network, based on fake news cascades, for rumor detection. Liu and Wu (2018) employ propagation path classification with RNN for early rumor detection. Zubiaga et al. (2018b) propose a method based on an LSTM layer followed by several dense ReLU layers. Other studies utilizing propagation path are (Kwon et al., 2017; Wu et al., 2015; Chen et al., 2016; Yang et al., 2018; Li et al., 2019a; Li et al., 2019b). Experiments from these studies show that models employing propagation path perform better than the feature-based algorithms. But we should keep in mind that we usually do not have much propagation information at the early stage of a rumor spread, and early detection is especially critical for a real-time rumor detection system. The study from (Vosoughi et al., 2018) shows that unconfirmed news tends to exhibit multiple and periodic discussion spikes, whereas confirmed news typically has a single prominent spike, and false rumor spreads farther, faster, and more widely than true news.

**Network-based:** Network-based rumor detection constructs flexible networks to indirectly capture rumor propagation information. The constructed networks can be homogeneous, heterogeneous, or hierarchical. Gupta et al. (2012) construct a network consisting of users, messages and events, using PageRank-like algorithm to compute event credibility. Yang et al. (2015) incorporate network features derived from comments, and they said that when the network feature was added to the traditional features, the results improved substantially. Wang and Terano (2015) propose social graphs to model the interaction between users and identify influential rumor spreaders. Heterogeneous networks have multiple types of nodes or edges. An example is the tri-relationship network among news creators, the rumors, and users (Shu et al., 2017b), which uses entity embedding and relation modeling to build a hybrid framework for rumor detection. In (Zhang et al., 2018), an RNN model is designed to detect rumors through exploring creators, contents, subjects and their relationships.

## 3.4 Joint Learning for User Stance and Rumor Detection

User stance plays an important role in rumor detection. Recent works have employed multi-task learning approaches to jointly learn stance detection and veracity prediction, in order to improve classification accuracy by utilizing the interdependence between them. Ma et al. (2018a) jointly learn the stance detection and the veracity prediction tasks, where each task has a task-specific GRU layer, and the tasks also share a GRU layer. The shared layer is to capture patterns common to both tasks, and the task specific layer is to capture the patterns that are more important to that task. In the rumor detection task, the hidden state at the last time step is used for prediction through a fully-connected output layer.

Ma et al. found that joint learning improves the performance of individual tasks, and utilizing shared and task-specific parameters is more beneficial than using only the shared parameters without the task-specific layer. Kochkina et al. (2018) propose a multi-task method without task specific layer for rumor verification. Both approaches do not employ attention in their models, and user information is not used. Li et al. (2019b) exploit both user credibility information and attention mechanism in their joint learning approach.

### 3.5 Rumor Detection Contests

There are two contests for rumor detection: 1. SemEval-2017 Task 8: Determining rumor veracity and support for rumors (Derczynski et al., 2017). The approach from (Enayet and El-Beltagy, 2017) was ranked No. 1 for the rumor detection task. 2. SemEval-2019 Task 7: Determining rumor veracity and support for rumors (Gorrell et al., 2019). The approach from (Li et al., 2019a) was ranked No. 1 for the rumor detection task. The datasets used in these two tasks are listed in Table 1. Both (Enayet and El-Beltagy, 2017) and (Li et al., 2019a) exploited content, user and propagation information. They also utilized user stance directly in their models. The main difference between them are that Li et al. (2019a) used neural networks, while Enayet and El-Beltagy (2017) employed an SVM model.

There are also two contests related to fake news, but actually both of them are about stance detection, not fake news detection. They are the *Fake News Challenge* at: http://www.fakenewschallenge.org, and the *WSDM 2019 cup: classification of fake news article* at: https://www.kaggle.com/c/fake-news-pair-classification-challenge

## 4 Future Research Directions

Although significant advances have been made in debunking rumors on social media, nevertheless, there remain many challenges to overcome. Based on the review of previous studies and also our experiences in both research and practical system implementation of rumor detection, here we present several directions for future rumor detection research.

### 4.1 Knowledge Base

Knowledge base (KB) is very helpful for fake news detection (Hassan et al., 2017). There have been some studies on employing KB for fake news detection, but very few or none on rumor detection over social media. One reason is that for rumors on social media, we already have much information, especially the social context information, to exploit and do research on. Another reason is that, compared to fake news detection which mainly deals with news articles, rumors on social media are about various topics, and it is hard to build appropriate KBs that cover them. Therefore, most previous studies on rumor detection have not paid attention to exploiting KB for debunking rumors.

The automatic fact-checking process aims to assess the claim by comparing the knowledge extracted from rumor text to known facts (true knowledge) stored in the constructed KB or knowledge graph. One advantage of utilizing KB for debunking rumor on social media is that the source posts (claims) are usually short, and it is easier to extract the main claim from the short message, compared to analyzing a long news article which might have several claims. Research from (Kwon et al., 2017) shows that text features are very important when we want to detect rumor at its very early stage, since there is no propagation information or very few feedbacks from users when a rumor just emerges. By extracting knowledge from rumor text, we hypothesize that the KB-based approach would be especially helpful for the rumor early detection. As a starting point, the initial research effort can focus on the topic areas of popular rumors, and the approaches that are already effective in fake news detection can be explored first. We think how effective KBs can help in rumor detection and how we can integrate it with other social context information will be an interesting research topic.

### 4.2 Target of User Response

User responses are quite informative for rumor detection. Usually false rumors will receive more negative and questioning responses, which can be leveraged for rumor detection. Each source message (rumor claim) has many replies, and they are either direct replies, or replies to other messages in the conversion thread. The structure of the conversion thread is important for understanding the real stance of the user of a reply. For example, given a message "This is fake" and a reply to it "I totally agree", if we do not consider that the reply is towards "This is fake", then we will give a wrong stance label, "support", to this reply. But actually, this response is denying the rumor claim. Although the neural network models based on propagation analysis may partially learn this information, we think explicitly handle this situation would improve rumor detection performance.

Another issue with the user response target is that sometimes the user response is not towards the claim of the source message, but certain aspects of the rumor story. For example, this is a false rumor in SemEval19 rumor detection task: "National Geographic channel has reportedly paid $ 1 million for this daring video. https://t.co/CDbjf65bKG." Many responses towards this rumor are talking about how great the video is or how brave the goat in the video is, e.g. "Perseverance and fighting spirit!!" and "Nice one!!!!!!". For a stance detection algorithm, it is very possible to predict this type of responses as "support", due to their positive sentiment. This obviously will also mislead the rumor detection algorithm. We think it is worthwhile to research on the intent of user responses, to better understand the actual target of a user comment.

### 4.3 Cross-domain and Cross-language

Most previous studies emphasize on distinguishing false rumors from truth with experimental settings that are generally limited to a specific social media platform, or certain topic domains, such as politics. Analyzing rumors across topics or platforms would let us gain deeper understanding of rumors and discover their unique characteristics that can further assist debunking them across domains (topic and platform).

Recently, we have seen rumors spreading across languages, especially rumors involving topics on politics, investment, business and finance. Often times, a rumor is already debunked in one language, but it is still spreading in another language, due to the language barrier and the lack of cross-language rumor detection tool. This is quite true for some rumors in Chinese on Weibo and WeChat, a social media platform similar to Facebook. These rumors are usually about politics, world affairs, business and health/medical topics.

For example, in WeChat, there are many rumors about some supplements, claiming they are good for certain diseases and also presenting certain fake evidences citing some foreign studies. This type of rumors is very hard for ordinary users to verify, especially the elder people who are the main group who are interested in rumors related to healthcare, medicine, and longevity. This has becoming more serious in the last couple of years, since more people in the rural areas start to use smart phone and social media. How to deal with this type of cross-language and cross-platform rumor detection problem would also be an interesting research topic.

### 4.4 Explanatory Detection

Most rumor detection approaches only predict the veracity of a rumor, and very little information is revealed why it is a false rumor. Finding the evidences supporting the prediction and presenting them to users would be very beneficial, since it helps users to debunk rumors by themselves. Making the result explanatory has attracted research in other areas, such as explanatory recommendation, but it is still a new topic in rumor detection field. This may become harder as more models are using deep learning techniques nowadays. However, as AI techniques are used in more applications, the demands for result explanation from users are also increasing. For example, now we are designing and implementing a rumor detection system for an Alibaba product, and one important product feature required by the product designers and users is to provide explanation for the veracity prediction result.

### 4.5 Integrating User Stance and User Credibility Together

Several studies have shown that both user stance and user credibility information help improve rumor detection performance (Liu et al., 2015; Enayet and El-Beltagy, 2017; Li et al., 2019b). However, these studies just treat the stance label and the features reflecting user credibility, such as no. of followers and user account age, as separate features in the overall prediction model. None of them has tried to integrate these two types of information together systematically, to get a unified indicator to reflect how important a response is for determining the veracity of a rumor. For example, we want to clearly differentiate these two different situations: an authoritative and credible user, such as a credible news agency or government agent, debunks or supports a claim, and a low credible user, e.g. a malicious account, debunks or supports a claim. And as explained in the "Target of User Response" section, we also need to take the real target into consideration when designing the integration model.

### 4.6 Utilizing External Textual Information

Besides KBs mentioned before, other types of external information may also help rumor detection, such as articles from credible new agency websites, announcements or documents from governments and authorities, official announcements from involved parties, past rumors that have been verified, etc. We can compare the current rumor with these external text data, to gain more insights on the rumor. This sounds like a boring idea and an old information retrieval and text matching problem, but actually it will have very practical impact on rumor detection, especially for a real rumor detection system. Many rumors are just resurfacing of old ones, or their variants. And for a human, when we verify a rumor, one of the things we will do is also to check relevant website to see if there is any relevant information about this rumor, such as official announcement. The study from (Qin et al. 2016) shows that this approach is very effective when detecting rumors that have variants in the past at real-time. One system implementation challenge is to monitor these websites and scrape the relevant text information.

### 4.7 Multi-task Learning

Studies already show that jointly learning of stance detection and rumor detection improves the performance of rumor detection (Kochkina et al., 2018; Ma et al., 2018a). In the rumor detection workflow, depending on the algorithms, the following tasks might be involved: user credibility evaluation, source credibility evaluation, knowledge extraction, etc. If there are appropriate datasets with annotations for these data types, one research direction is to explore multi-task learning for these tasks, in addition to the user stance and rumor detection tasks. We expect it will benefit the rumor veracity prediction task, at least.

### 4.8 Rumor Early Detection

Rumor early detection is to detect a rumor at its early stage before it wide-spreads on social media, so that one can take appropriate actions earlier. Early detection is especially important for a real-time system, since the more a rumor spreads, the more damages it causes, and more likely for people to trust it. This is a very challenging task, since at its early stage a rumor has little propagation information and very few user responses. The algorithm has to primarily rely on the content and external knowledge, such as KB. Several studies have tested their algorithms on the early stage of rumors (Liu et al., 2015, Ma et al., 2016; Kwon et al., 2017; Liu and Wu, 2018). Kwon et al. (2017) analyzed feature stability over time and reported that user and linguistic features are better than structured and propagation features for determining the veracity of a rumor at its early stage. Although there are already some studies on this direction, more research efforts are still needed, due to its importance in the real systems.

### 4.9 Framework for a Real Rumor Detection System

Although there are many studies on rumor detection, most of them focus on models that utilize only part of the available information and test them on datasets that are platform or domain-specific. Very few of them are designed for real-time systems (Liu et al., 2015; Liu et al., 2016). A framework for a practical rumor detection system should try to exploit all the available information, and apply these information and models appropriately for different situations that might involve multiple factors, such as platforms, rumor stages, topics, languages, and content types (text, video or image). From the exploiting information point of view, we think the following information or data are worth to explore: text content (lexical, syntactical, semantic, writing style, etc.), visual content (video, image), rumor topics, knowledge bases, external documents, old rumors, propagation information, user features, source credibility, user credibility, heterogenous and homogeneous network structures, cross-platform information, and cross-language information.